# Coupled plasmonic Fabry–Pérot resonances modulating electromagnetic enhancement at a silver nanowire–nanoparticle junction proved by surface-enhanced fluorescence spectroscopy

Tamitake Itoh[1,a)] and Yuko S. Yamamoto[2]

[1]Health and Medical Research Institute, National Institute of Advanced Industrial Science and Technology (AIST), Takamatsu, Kagawa 761-0395, Japan

[2]Graduate School of Advanced Science and Technology, Japan Advanced Institute of Science and Technology (JAIST), Nomi, Ishikawa 923-1292, Japan

[a)]Author to whom correspondence should be addressed: tamitake-itou@aist.go.jp

**ABSTRACT**

We investigated the surface-enhanced fluorescence (SEF), which is the broad background spectrum underlying the surface-enhanced resonant Raman scattering of dye molecules, emitted from junctions between silver nanoparticles (NPs) and nanowires (NWs). We found that the electromagnetic enhancement factors ($F_R$) derived from the SEF exhibited spectral modulation, such as complicated oscillatory structures. This property became clearer when the emission polarization was perpendicular to the long axis of the NWs. The $F_R$ spectra of the long and short NWs exhibit fine and broad oscillatory structures, respectively, when the NPs are positioned near the NW centers. When the NPs were near the NW ends, a combination of fine and broad structures was observed. Numerical calculations of the position-dependent $F_R$ spectra successfully reproduced these properties, indicating that the spectral properties originated from the interaction between two plasmonic Fabry–Pérot (FP) resonators within the NW, separated by a NP. Based on this result, an analytical model of coupled FP resonators was developed to evaluate these spectral features using experimentally measured NW lengths and NP positions while varying the inter-resonator coupling efficiency. This analysis revealed that the interference and superposition of the two FP resonances resulted in these complex spectra, the energy of which was radiated by the localized plasmons of the NPs.

## I. INTRODUCTION

A spontaneous de-excitation rate of a molecule interacting with localized plasmons (LPs) of a gold or silver nanoparticle (NP) is enhanced by Purcell factor $F = [Q(\lambda/n)^3]/(4\pi^2 V_{\mathrm{LP}})$, where $Q$ is the quality factor of the LP resonance; $\lambda$ and $n$ are the light wavelength and refractive index around the NP, respectively; and $V_{\mathrm{LP}}$ is the mode volume of the LP resonance (assuming that the direction of the dipole is random, $F = [3Q(\lambda/n)^3]/(4\pi^2 V_{\mathrm{LP}})$).[1] A large electromagnetic (EM) enhancement factor ($F_{\mathrm{R}}$), which is a radiative portion of $F$, of NP aggregates such as a NP dimer has reached to $10^5$,[2–4] enabling single molecule Raman spectroscopy with two-fold EM enhancement.[5–8] The large $F_{\mathrm{R}}$ has been applied to various surface-enhanced (SE) spectroscopies, including SE-Raman (resonant) scattering [SE(R)RS], SE-fluorescence (SEF), SE-absorption, and their nonlinear counterparts, such as SE-hyper-Raman scattering (SEHRS).[9] The origin of this large $F_{\mathrm{R}}$ is the extremely small $V_{\mathrm{LP}}$ of coupled LPs between two NPs separated by a nanogap or junction.[1–4,9] Such location with large $F_{\mathrm{R}}$ is referred to as a hotspot (HS). These HSs are also a key to realizing exotic phenomena related to cavity quantum electrodynamics (cavity-QED), such as strong-to-ultrastrong coupling, molecular optomechanics, and polariton chemistry.[10–16] Gold or silver nanowires (NWs) have been used for various SE spectroscopies and cavity-QED because surface plasmon polaritons (SPPs) propagating along NWs enable remote excitation and detection at the NW ends.[17–19] At NW ends, SPPs are converted into light by overcoming the momentum mismatch between free-space light and SPPs.[20] The HS composed of a NW and a NP have also been utilized for unique plasmonic applications including remote SE spectroscopy, remote excitation of optical activity, chiral light generation and sensing, unidirectional light emission, strong coupling, and Fano resonance.[21–29] The $F_R$ of this type of HS is generated by two types of coupled plasmons— one between LPs of a NW and LPs of a NP [Fig. 1(a)], and the other between surface plasmons (SPs) of a NW and LPs of a NP [Fig. 1(b)].[30] The latter coupled plasmons exhibit the broad $F_R$ spectral width at the HS from the visible to the near-infrared (NIR) region because the SP, whose energy spectrum extends continuously from the ultraviolet region to zero, can couple with both quadrupole LP (visible region) and dipole LP (NIR region) of NPs.[30–32] Such spectrally broad $F_R$ at the HSs enables SEHRS—which involves NIR excitation and visible emission—even under continuous-wave excitation.[30,33]

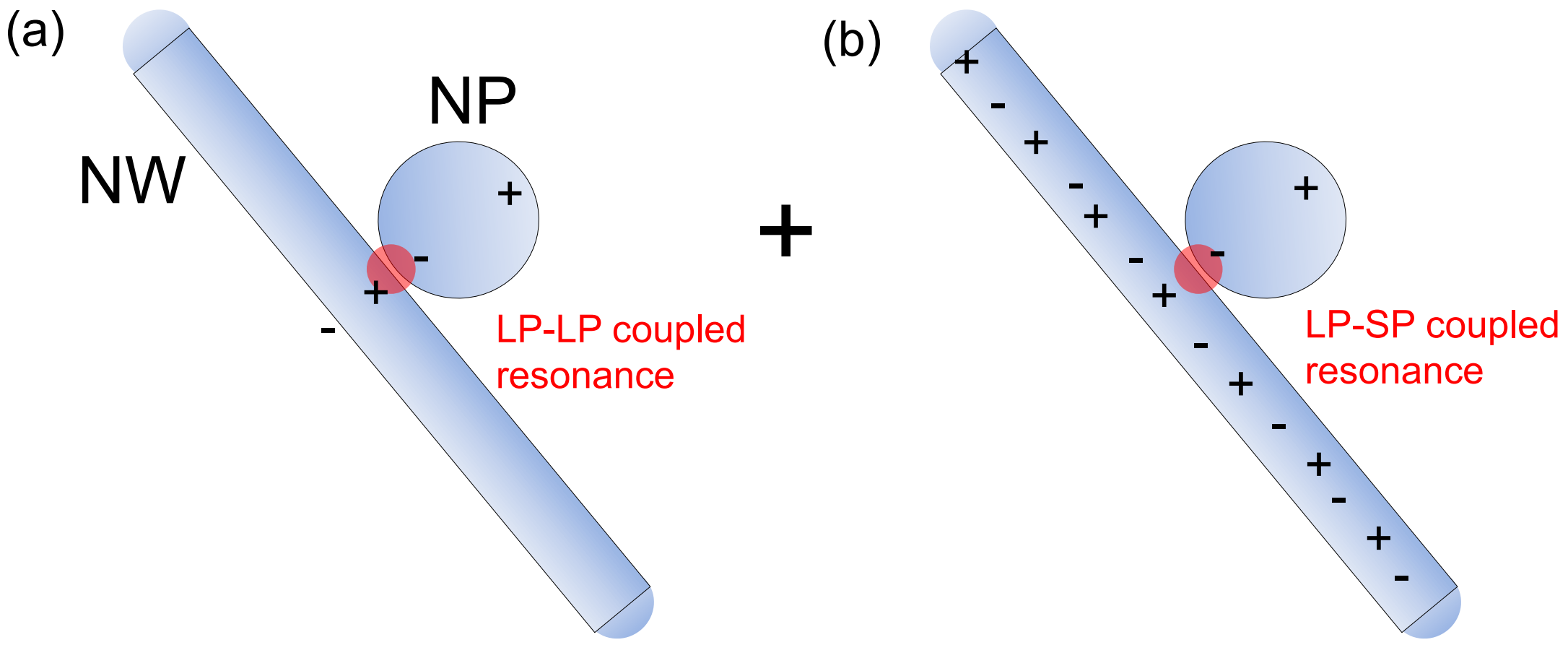

FIG. 1. Schematic images of (a) a coupled plasmon between a localized plasmon (LP) of nanoparticle (NP) and an LP of nanowire (NW) and (b) a coupled plasmon between an LP of NP and a surface plasmon (SP) of NW.

However, the application of this type of HS to SE spectroscopy is an important issue that needs to be addressed. An NW with a finite length exhibits Fabry–Pérot (FP) resonance, which arises from multiple back-and-forth reflections of SPPs between the two ends of the NW. Multiple reflections result in oscillatory structures in the Rayleigh scattering spectra detected from the NW ends.[18,19,34] Thus, $F_R$ at the HS between a NP and a NW may be considerably affected by the FP resonance. Addressing this issue is crucial for accurately analyzing the SE spectroscopic and cavity-QED results. The spectral properties of the FP resonance of NWs have mainly been studied using the Rayleigh scattering cross-section ($\sigma_{sca}$) from both ends of the NWs.[18,19,34] However, the $\sigma_{sca}$ spectrum includes the background from HS surrounding because its detection area cannot exceed the optical diffraction limit. Furthermore, considerable deviation between the $\sigma_{\mathrm{sca}}$ and $F_{\mathrm{R}}$ spectra occurs when a subradiant plasmon like dipole-quadrupole coupled mode is involved in the $F_{\mathrm{R}}$.[35,36] Indeed, the quantum optical origins of $\sigma_{\mathrm{sca}}$ and $F_{\mathrm{R}}$ correspond to the $\hat{A}^2$ and $\hat{A} \cdot \hat{p}$ terms, respectively, where $\hat{A}$ and $\hat{p}$ are the vector potential and momentum operators.[37–41] Therefore, a method for directly evaluating the effect of FP resonance on the $F_{\mathrm{R}}$ spectrum at the HS needs to be developed.

In this study, the effect of the FP resonance on the HSs between silver NPs and NWs was investigated using both $\sigma_{\mathrm{sca}}$ and $F_{\mathrm{R}}$ spectra. The $F_{\mathrm{R}}$ spectra were derived using the SEF spectra, which appear as a broad background of the SERRS spectra and are described by a two-step $\hat{A} \cdot \hat{p}$ process involving one-photon excitation and one-photon emission.[39–41] Both $\sigma_{\mathrm{sca}}$ and $F_{\mathrm{R}}$ spectra clearly exhibit complicated oscillatory structures when the emission polarization is perpendicular to the long axis of the NWs. These spectral structures depend considerably on both the NW length and the position of the NP on the NW. Numerical calculations of the $F_{\mathrm{R}}$ spectra and spatial distributions of $F_{\mathrm{R}}$ along the NWs indicated that these spectral structures arose from the interaction between the two FP resonators separated by the NP on the NW. Based on this indication, we developed an analytical model of coupled FP resonators to evaluate these spectral structures as a function of the NW length, NP position, and inter-resonator coupling efficiencies. This analysis revealed that the interference and superposition of the SPPs of the two FP resonators resulted in the complex oscillatory structures observed in $\sigma_{sca}$ and $F_R$ spectra.

## II. METHODS

The silver NWs colloidal dispersion was obtained from University of Shiga Prefecture; the preparation method is described in Ref. 42. Silver NPs are by-products of the NW colloidal dispersion. The average NW radius and length were approximately 35 nm and 10 μm, respectively, and the distribution of the NP radius was approximately 100 nm.[30] A mixture of the dispersion of NWs, including NPs and crystal violet (CV) methanol solution ($5.0 \times 10^{-6}$ M) was dropped and heat-dried on a glass plate. We considered that the CV molecules accessed the HSs between NPs and NWs during the drying process. The effective concentration of the dye on the sample was reduced by photobleaching most of the excess dye molecules adsorbed onto the NWs, NPs, and glass surfaces via 532-nm wavelength laser beam irradiation for 3 min. Photobleaching was confirmed by the disappearance of the fluorescence from the glass and isolated NW surfaces.

The Rayleigh scattering intensity spectra from the HSs between the NPs and NWs were acquired by illuminating white light from a 50 W halogen lamp using a dark-field condenser with a numerical aperture (NA) of 0.92. Rayleigh scattered light was collected using an objective (UPlanFL N, Olympus, Japan) with an NA of 0.6. The SEF (including SERRS) intensity spectra of identical HSs were measured by illumination with an unpolarized green laser beam (532 nm, 35 W/cm$^2$; CW DPSS 532, Laser Create, Japan) using a lens (NA = 0.2). The SEF light was collected using a common objective by changing the NA from 0.92 to 1.3. The excitation wavelength was shorter than that of the fluorescence peak (636 nm). The shorter excitation wavelength was used to expand the spectral window of $F_R(\lambda)$ spectra.[43] Rayleigh scattering and SEF light from single HS located at the center of the microscopic image were selected using a pinhole (diameter of 3 μm on the sample surface) and directed to a polychromator equipped with a thermoelectrically cooled charge-coupled device assembly with a grating groove-density of 150 grooves/mm (DV 437-OE-MCI, Andor, Japan). Figures 2(a1) and 2(a2) show a scanning electron microscopy (SEM) micrograph of a NP on the NW and its magnified image, respectively. Figures 2(b) and 2(c) show the dark-field and SEF images of the NP [Fig. 2(a)], respectively. The SEF image confirmed the formation of HS at the junction between the NP and NW.

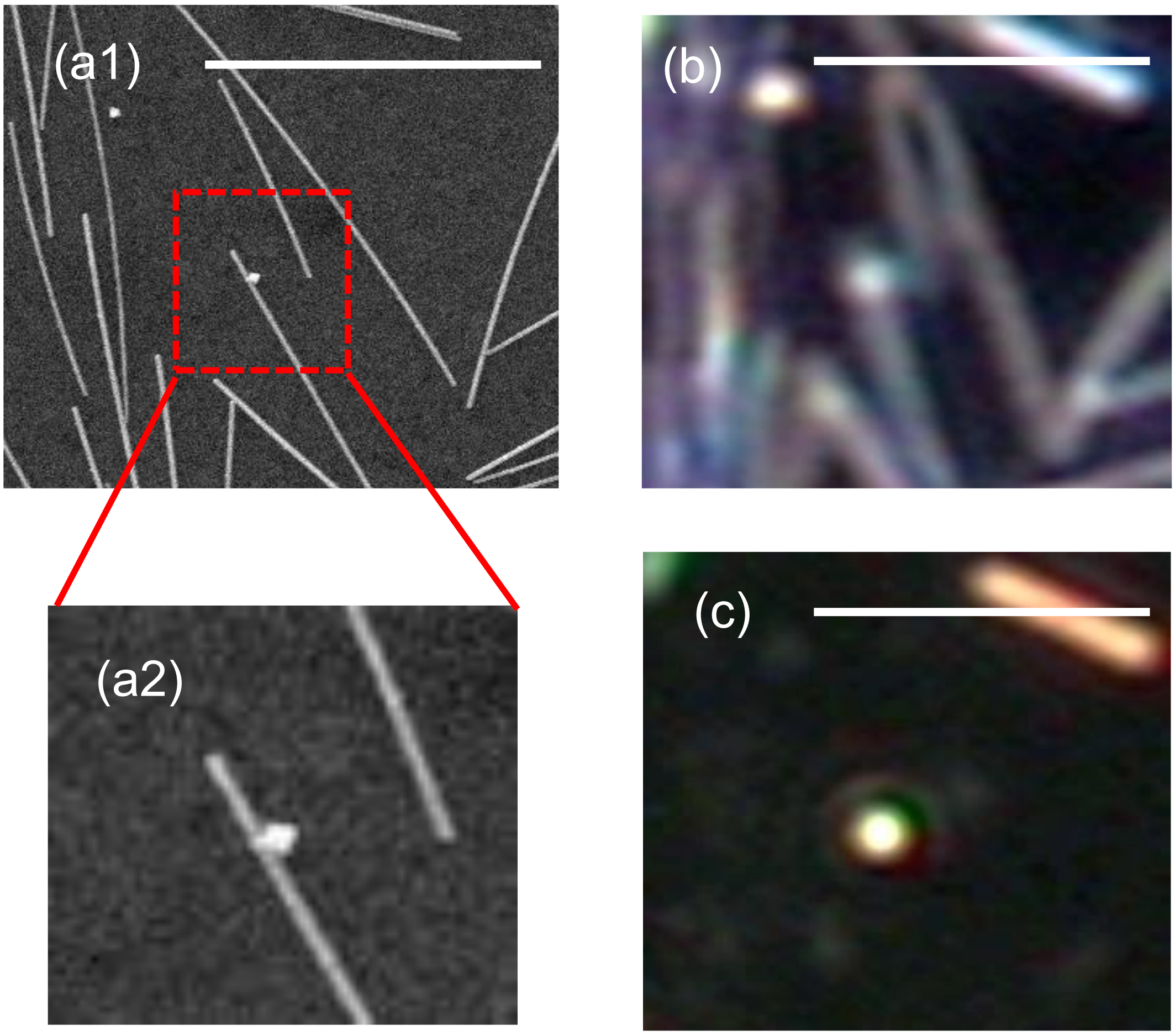

FIG. 2. (a1) SEM micrograph and (a2) a magnified image of the hotspot composed of a NP and a NW. (b) Dark-field and (c) SEF images of HS. All scale bars represent 4 μm.

## III. RESULTS AND DISCUSSION

The procedures for obtaining $\sigma_{sca}(\lambda)$ and $F_R(\lambda)$ spectra is as follows. Figures 3(a1)–3(a3) illustrate the procedure for converting the Rayleigh scattering intensity spectra of a HS into $\sigma_{sca}(\lambda)$ spectra. The effect of the white-light spectrum on the Rayleigh scattering intensity spectrum of the HS was eliminated by dividing the scattering intensity spectrum by the white-light spectrum, as shown in Figs. 3(a1) and 3(a2), resulting in conversion to the $\sigma_{sca}(\lambda)$ spectra [Fig. 3(a3)].

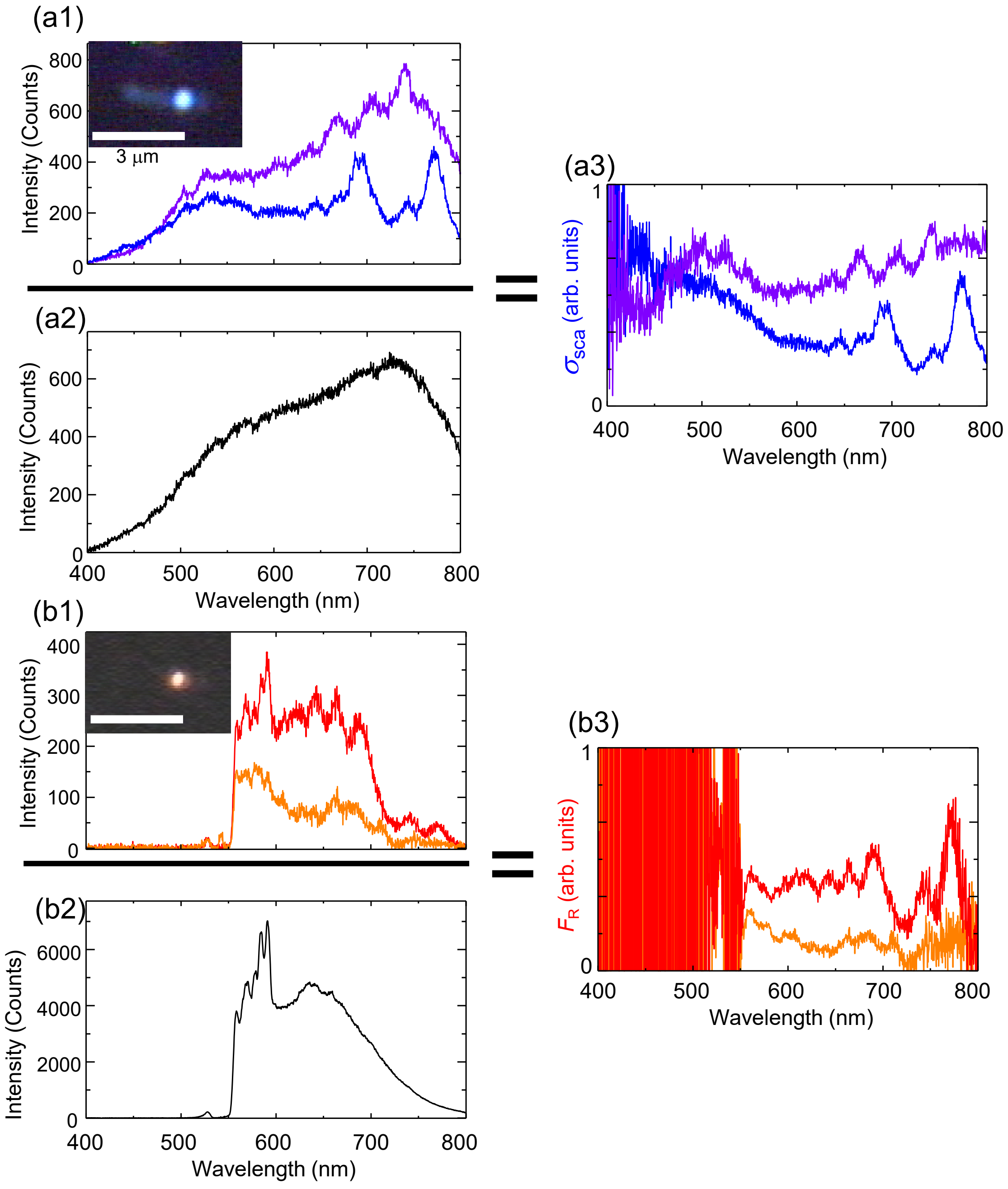

FIG. 3. Procedures for obtaining $\sigma_{sca}(\lambda)$ and $F_R(\lambda)$. (a1) Typical Rayleigh scattering intensity spectra of a HS with emission polarization perpendicular (blue curve) and parallel (purple curve) to the long axis of NW. Inset shows a dark-field image of the HS between a NP and a NW. (a2) Spectrum of the white light source. (a3) $\sigma_{sca}(\lambda)$ spectra with emission polarization perpendicular

(blue curve) and parallel (purple curve) to the long axis of NW obtained by dividing the Rayleigh scattering intensity spectra by the white light spectrum. (b1) Typical SEF (including SERRS) spectra of the HS with emission polarization perpendicular (red curve) and parallel (orange curve) to the long axis of NW with excitation laser energy of 2.33 eV (532 nm). The laser line was omitted by the notch filters. SERRS lines are merged because of low groove density grating. Inset shows an SEF image of the HS. (b2) SEF spectrum (including SERRS) of large silver NP aggregates. (b3) $F_R(\lambda)$ spectra with emission polarization perpendicular (red curve) and parallel (orange curve) to the long axis of NW obtained by dividing the SEF spectra by the SEF spectrum of the large aggregates. The scale bars are 3 μm.

We examined the $\sigma_{sca}(\lambda)$ spectra with two scattering light polarizations, parallel and perpendicular to the long axis of the NW. The complex spectral structures appear more clearly for perpendicular polarization, indicating efficient coupling between the SPs of the NW and LPs of the NP in this polarization direction.[30–32] Figures 3(b1)–3(b3) illustrate the procedure for converting the SEF (including SERRS) intensity spectra of the HS to respective $F_R(\lambda)$ spectra. The $F_R(\lambda)$ spectrum of the large NP aggregates [Fig. 3(b2)] is considered flat because it is the sum of various $F_R(\omega)$ spectra from a large number of HSs.[39,40,43] Thus, the $F_R(\lambda)$ spectra in Fig. 3(b3) were obtained by dividing the SEF intensity spectra in Fig. 3(b1) by that of the large NP aggregates in Fig. 3(b2). We also examined the $F_R(\lambda)$ spectra with two emission light polarizations, parallel and perpendicular to the long axis of the NW. The complex spectral structures also appeared more clearly for perpendicular polarization, indicating efficient coupling between the SPs of the NW and the LPs of the NP.[30–32] Because of the clearer appearance of complicated spectral structures in $\sigma_{sca}(\lambda)$ and $F_R(\lambda)$ spectra for the polarization perpendicular to the long axis of the NW, spectral detections were performed in this polarization direction.

We demonstrate two typical examples of the $\sigma_{sca}(\lambda)$ and $F_R(\lambda)$ spectra, which show complex oscillatory structures. One is the $\sigma_{sca}(\lambda)$ and $F_R(\lambda)$ spectra of the HSs, where the NPs are located near the centers of the NWs, and the other is for the HSs, where the NPs are located near the ends of the NWs. Figures 4(a1-i)–4(a4-i) and 4(a1-ii)–4(a4-ii) show the four dark-field images and $\sigma_{sca}(\lambda)$ with $F_R(\lambda)$ spectra, respectively, for the HSs where the NPs are located near the centers of the NWs. Both the $\sigma_{sca}(\lambda)$ and $F_R(\lambda)$ spectra show that the oscillatory periods became shorter as the NW length of NWs become longer. Such tendency is consistent with the properties of FP resonances of isolated NWs by considering NW length dependence of constructive and destructive interference conditions of SPPs.[34,44,45] The degree of similarity in peak positions between $\sigma_{sca}(\lambda)$ with $F_R(\lambda)$ spectra varies depending on HSs, indicating the different effect of $\hat{A}^2$and $\hat{A} \cdot \hat{p}$ terms on the $\sigma_{sca}(\lambda)$ and $F_R(\lambda)$, respectively.[40,41,43] Indeed, the peaks in $F_R(\lambda)$ appear in slightly shorter wavelength regions than those in $\sigma_{sca}(\lambda)$.[40,41,43] Figures 4(b1-i)–4(b4-i) and 4(b1-ii)–4(b4-ii) are the four dark-field images and $\sigma_{sca}(\lambda)$ and $F_R(\lambda)$ spectra, respectively, for the HSs where NPs located near the ends of NWs. Neither the $\sigma_{sca}(\lambda)$ and the $F_{\mathrm{R}}(\lambda)$ spectra clearly show periods in the oscillatory structures; instead, they exhibit complicated structures composed of large broad peaks and small narrow peaks, indicating that the two types of FP resonances overlap in the spectra. The degree of similarity in the peak positions between the $\sigma_{\mathrm{sca}}(\lambda)$ and $F_{\mathrm{R}}(\lambda)$ spectra also varies depending on the HSs, indicating a difference between the $\hat{A}^2$ and $\hat{A}\cdot\hat{p}$ terms.[40,41,43]

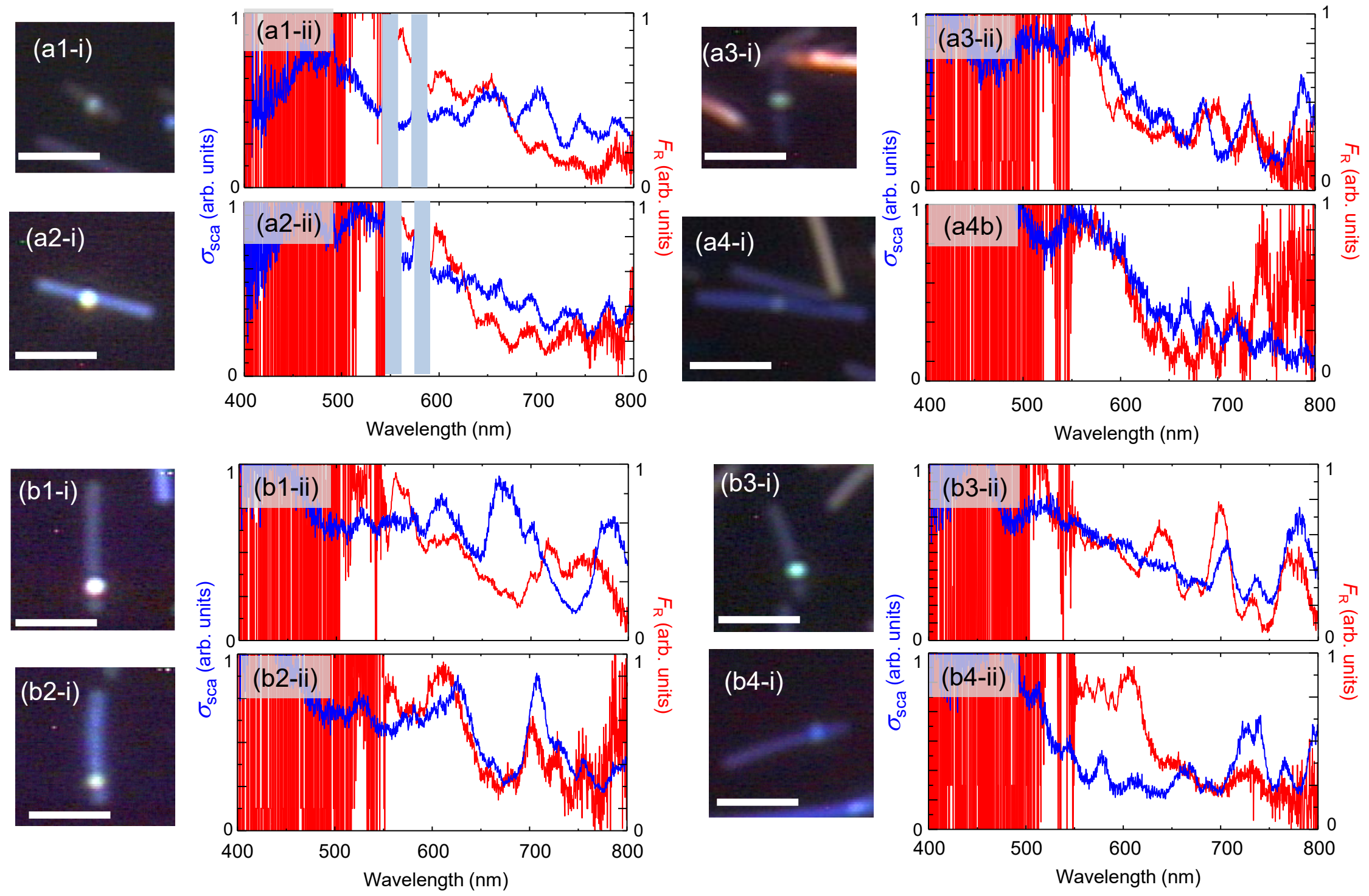

FIG. 4. (a1-i)–(a4-i) and (a1-ii)–(a4-ii) Four data sets of dark-field images and spectra of $\sigma_{\mathrm{sca}}(\lambda)$ (blue curves) and $F_{\mathrm{R}}(\lambda)$ (red curves), respectively, for the HSs where NPs located near the centers of NWs. (b1-i)–(b4-i) and (b1-ii)–(b4-ii) Four data sets of dark-field images and spectra of $\sigma_{\mathrm{sca}}(\lambda)$ (blue curves) and $F_{\mathrm{R}}(\lambda)$ (red curves), respectively, for the HSs where NPs located near the ends of NWs. All scale bars represent 3 μm.

Both Figs. 4(a) and 4(b) show the oscillatory structures in the region $\lambda > 600$ nm.

Alternatively, large broad structures appeared in the region $\lambda < 600$ nm, indicating that these broad structures were due to the coupled plasmon resonances between the LPs of the NPs and NWs. Rayleigh scattering of the coupled LP resonance significantly attenuated the amplitudes of the SPPs when passing through the NPs, resulting in the disappearance of the oscillatory structures in the region $\lambda > 600$ nm. Both Figs. 4(a) and 4(b) show the HS-by-HS variations in the depths of the complex oscillatory structures. In the case of isolated NWs, such depths are determined by the reflection coefficient at the ends of the NWs and the length of the NWs.[34] The depths become deeper as the reflection coefficient increases because of the increase in the number of multiple back-and-forth reflections, enhancing the constructive and destructive interference of SPPs. However, the reflection coefficients of the NW ends are almost identical. Thus, the current variations may be determined by Rayleigh scattering at the HSs. As the NW lengths increase, the depths become shallower owing to the larger ohmic loss, which limits the number of reflections. Thus, the variation may also be determined by the NW length.

We performed numerical calculations to elucidate the above-described spectral properties in $\sigma_{sca}(\lambda)$ and $F_R(\lambda)$ at the HSs [Figs. 4(a) and 4(b)] using the Finite-Difference Time-Domain (FDTD) method (EEM-FDM Version 5.1, EEM Co., Ltd., Japan). The complex refractive index of the silver NPs and NWs was adopted from a previous report.[46]

The effective refractive index of the surrounding medium was set to 1.39 to ensure consistency between the calculated and experimental $\sigma_{sca}(\lambda)$ spectra of the gold NPs on a glass plate.[38] In these calculations, quantum tunnelling and nonlocal effects, both of which reduce $F_R$ by unscreened surface electrons,[2,3,47] were not considered. Thus, the calculated $F_R(\lambda)$ may have been overestimated. In FDTD calculation, we approximated the shapes of the NPs and NWs as spherical and cylindrical with hemispherical ends, respectively. However, in reality, the cross-section of NWs is pentagonal.[48] The purpose of the calculation was not to quantitatively reproduce the experimental results but rather to clarify the EM coupling between a LP of NP and a SP of NW, which generates complex oscillatory structures in the $\sigma_{sca}(\lambda)$ and $F_R(\lambda)$ spectra.

For FDTD calculations, we assumed that the FP resonances in $\sigma_{sca}(\lambda)$ and $F_R(\lambda)$ [Figs. 4(a) and 4(b)] were affected by the relative positions of the NPs on the NWs. Therefore, the position dependence was examined using calculations. The six NP positions on a NW used to reproduce the spectral properties in $\sigma_{sca}(\lambda)$ and $F_R(\lambda)$ in Figs. 4(a) and 4(b) are shown in Figs. 5(a1)–5(a6). The lengths of the left and right sides of the NW, separated by the junction between the NP and NW, were defined as $L_1$ and $L_2$, respectively. The total length of the NW, $L_1 + L_2$ is set to 6.0 μm. Figure 5(b) shows the illumination setup and the coordinate system. The beam diameter of the white-light excitation was set to 600 nm.

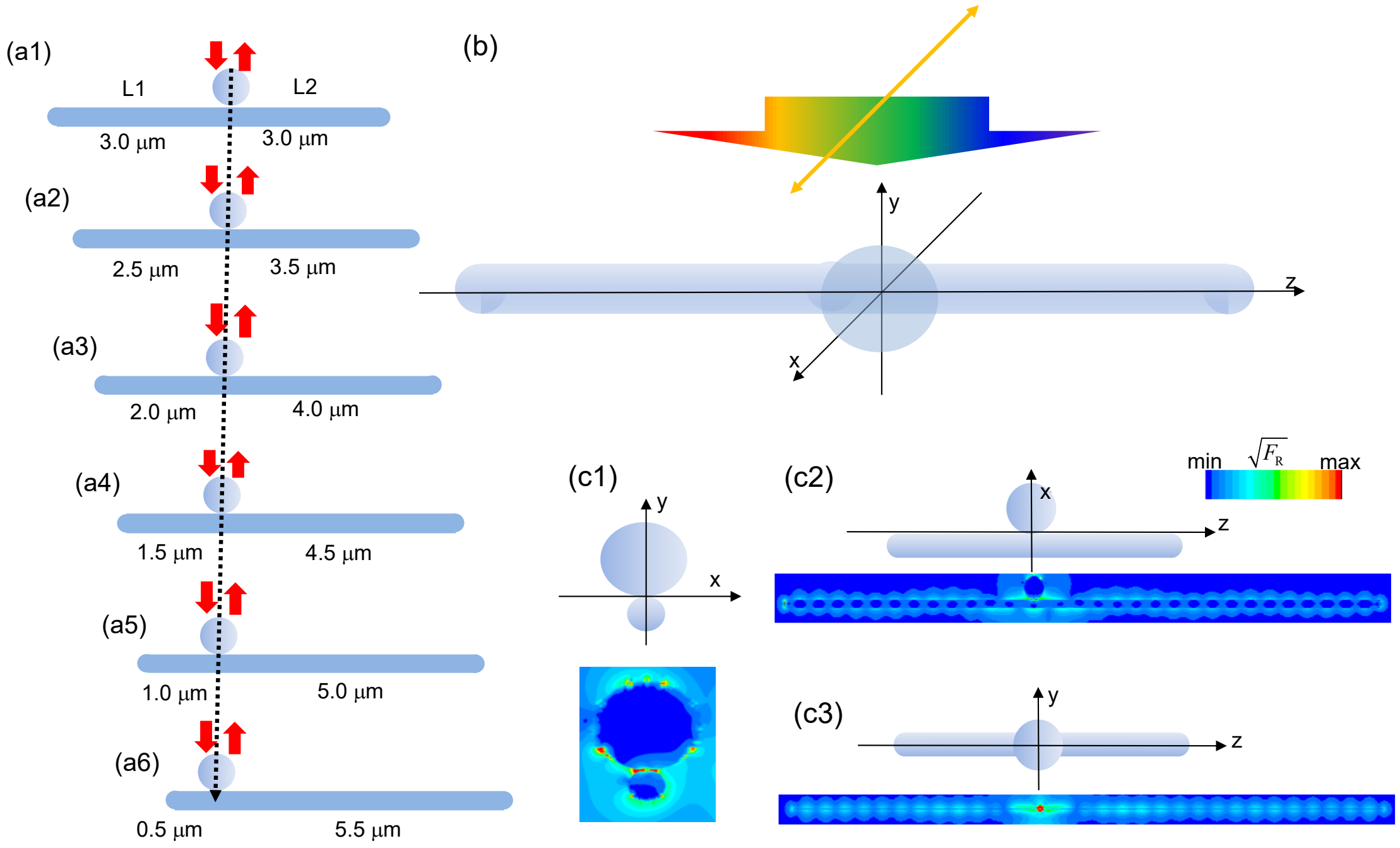

FIG. 5. (a1)–(a6) Schematic images of six positions of NPs on the NW. The values of $L_1$ and $L_2$ are indicated in each panel. $L_1 + L_2$ is set to 6 μm. (b) FDTD calculation setup for $\sqrt{F_R}$ around the system composed of an NP on an NW with gap $d_g = 0$ nm illuminated by a narrow-field white-light beam with a diameter of 600 nm from the upper side (y-axis). Polarization directions of white light are set to be perpendicular to the long axis of NW. (c1)–(c3) Schematic images of FDTD calculation setups for $\sqrt{F_R}$ distributions with perpendicular excitation around the system showing the (c1) x-y plane at $z = 0$ nm, (c2) x-z plane at $y = 0$ nm, and (c3) y-z plane at $x = 0$ nm for $\lambda = 525$ nm.

The excitation polarization direction was set perpendicular to the long axis of the NW because the complex properties of $\sigma_{sca}(\lambda)$ and $F_R(\lambda)$ became clear in this polarization direction, as shown in Fig. 3. The radius of NP ($R_{NP}$) and NW ($R_{NW}$) were set to 100 and 35 nm, respectively, based on the experimentally measured mean values.[30] The gap distance $d_g$ between the NP and NW was set to 0 nm, based on the SEM images of the HS [Figs. 2(b1) and 2(b2)] and the reported data.[22] The amplitude of incident electric field $|E_{in}|$ was set to 1.0 V/m. Thus, the calculated amplitude $|E_{loc}|$ corresponds to $|E_{loc}/E_{in}|= \sqrt{F_R}$. The $\sqrt{F_R}$ distributions at the junction and through the NW are shown in Fig. 5(c). Figures 5(c1)–5(c3) show the x-y ($z$ = 0), x-z ($y$ = 0), and y-z ($x$ = 0) plane images of $\sqrt{F_R(\lambda)}$ at $\lambda$ = 525 nm. A strong $\sqrt{F_R}$ around the junction indicates the presence of HS. The periodic structures of $\sqrt{F_R}$ appearing in the x-z and y-z plane images in Figs. 5(c2) and 5(c3) are the result of back-and-forth propagation of SPPs along the NW, starting from the HS. This back-and-forth propagation is the origin of FP resonances.[30,34]

We examined both spectral and spatial properties of the FP resonances by FDTD calculations. Figures 6(a1)–6(a6) show the six NP position dependences, corresponding to Figs. 5(a1)–5(a6), of spatial distributions of $\sqrt{F_R(\lambda)}$ along the NWs expressed as contour maps. These contour maps clearly exhibit the two oscillatory patterns on both sides of NPs induced by back-and-forth reflections of SPPs between the NW ends and the junction at $z$ = 0. These two oscillatory patterns largely differ from each other for the case in which the NP locates near the end of NW. These results indicate that two FP resonators with cavity lengths of $L_1$ and $L_2$ forms on both sides across the NPs. Figures 6(b1)–6(b6) show the magnified contour maps of Figs. 6(a1)–6(a6) around the NP positions. Both sides of NPs exhibit the large values of $\sqrt{F_{\mathrm{R}}(\lambda)}$ indicating HSs. The spectral structures of $\sqrt{F_{\mathrm{R}}(\lambda)}$ on both sides differ as the NP approaches the end of NW. However, both spectral structures of $\sqrt{F_{\mathrm{R}}(\lambda)}$ also show similarity, even for large difference between $L_1$ and $L_2$, as clearly shown in Fig. 6(a6). This coexistence of difference and similarity indicates the interaction between two FP resonators.

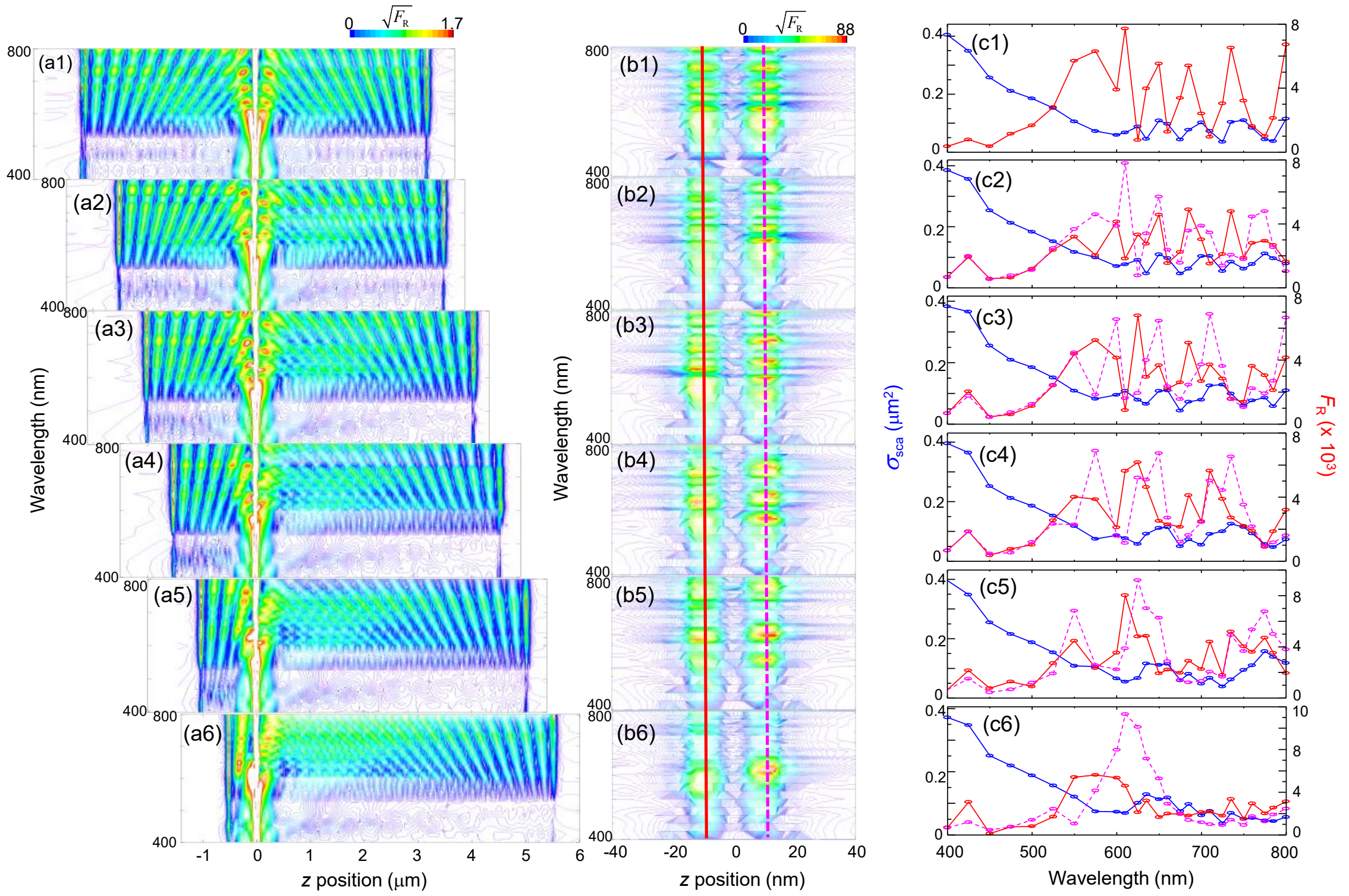

FIG. 6. (a1)–(a6) $\sqrt{F_{\mathrm{R}}(\lambda)}$ profiles along $z$-axis for $x$, $y$ = 0 nm, $R_{\mathrm{NP}}$ and $R_{\mathrm{NW}}$ of 100 and 35 nm, respectively, $d_{\mathrm{g}}$ = 0 nm, and $\lambda$ values of 400–800 nm with perpendicular excitation light expressed as contour maps. The values of $L_1$ and $L_2$ correspond to those of Figs. 5(a1)–5(a6). White areas around $z$ = 0 means $\sqrt{F_{\mathrm{R}}(\lambda)}$ > 1.7. (b1)–(b6) $\sqrt{F_{\mathrm{R}}}$ profiles along $z$-axis from −40 to 40 nm of (a1)–(a6), expressed as contour maps. The two maxima in $\sqrt{F_{\mathrm{R}}}$ spectra in $NW_1$ and $NW_2$ regions are traced by the red solid and pink dashed lines. (c1)–(c6) Spectra of $\sigma_{\mathrm{sca}}(\lambda)$ (solid blue) and two $F_{\mathrm{R}}(\lambda)$ curves for $NW_1$ region (solid red) and $NW_2$ region (dashed pink) with perpendicular excitation light. The values of solid red and dashed pink curves were derived from the red solid and pink dashed lines in (b1)–(b6).

Further, we extracted $F_R(\lambda)$ spectra at the two HSs, as indicated by the red solid and pink dashed lines in Figs. 6(b1)–6(b6) and compared them with $\sigma_{sca}(\lambda)$ spectra, as shown in Figs. 6(c1)–6(c6). These spectra exhibit both complex and oscillatory structures in the region $\lambda$ > 600 nm, supporting the experimental results in Fig. 4 and confirming the destruction of FP resonances by Rayleigh scattering of the coupled LP resonance at the junctions. When the NP is located at the center of the NW, as shown in Figs. 6(c1), both $\sigma_{sca}(\lambda)$ and two $F_R(\lambda)$ spectra exhibit similar oscillatory structures. The spectral peaks in $\sigma_{sca}(\lambda)$ are slightly redshifted relative to those in $F_R(\lambda)$ because $\sigma_{sca}(\lambda)$ and two $F_R(\lambda)$ originate from the $\hat{A}^2$and $\hat{A} \cdot \hat{p}$ terms, respectively, of quantum optics, supporting the discussion of Fig. 4.[39] This spectral similarity become unclear as the NP approaches the NW ends, as shown in Figs. 6(c2)–6(c4). Both $\sigma_{sca}(\lambda)$ and the two $F_R(\lambda)$ exhibit complicated structures, suggesting the overlap of the two FP resonance spectra with similar periods. Finally, both $F_R(\lambda)$ spectra exhibited the coexistence of large, broad, and weak fine peak structures, as shown in Figs. 6(c5) and 6(c6), suggesting the overlap of two FP resonance spectra with largely different periods. When the FP resonances were completely separated by the two NW regions, $NW_1$ and $NW_2$, $F_R(\lambda)$ spectra of the longer and shorter sides exhibited fine and broad structures, respectively. However, $F_R(\lambda)$ spectra show both fine and broad structures, as shown in Figs. 6(c5) and 6(c6), indicating an interaction between the two FP resonators via the HSs. The FDTD calculations revealed that the experimentally observed oscillatory and complicated spectral structures in $\sigma_{sca}(\lambda)$ and $F_R(\lambda)$ in Fig. 4 were determined by the difference between $L_1$ and $L_2$ of the FP resonators. The coexistence of broad and fine structures in both $F_R(\lambda)$ spectra indicates the interaction of the two FP resonances.

The FDTD calculations revealed that the experimental results of $\sigma_{sca}(\lambda)$ and $F_R(\lambda)$ can be reproduced by coupling the two FP resonators. Thus, we developed an analytical model of coupled FP resonators and evaluated their spectral properties using experimentally measured NW lengths and NP positions while varying the inter-resonator coupling efficiency. This analytical model was developed based on the conventional FP model.[49,50]

The characteristics of coupled FP resonators are determined by $L_1$ and $L_2$, the coupling efficiency at the junction, reflection coefficient at NW ends, and the complex propagation constant ($k_{spp} = \beta + i\alpha$) of SPP propagating a NW, where $\beta$ and $i\alpha$ are real and imaginary part of the constant, respectively. We first solved the dispersion relation of $k_{spp}$ for a NW based on previous reports.[50–52] The complex dielectric functions of relative permittivity of silver $\varepsilon_1$ were obtained from the experimental data reported by Johnson and Christy.[46] The NW was modeled as an infinitely-long cylinder with $R_{NW}$ of 35 nm embedded in a homogeneous dielectric medium with a relative permittivity of $\varepsilon_2$ (= $1.39^2$).[38] We focused on fundamental transverse magnetic ($TM_0$) mode, which is the lowest long-range mode and determines the characteristics of coupled FP resonators. The $TM_0$ mode satisfies the following characteristic eigenvalue equation.

$$\frac{\varepsilon_1(\lambda)}{\kappa_1}\frac{I_1(\kappa_1 R_{\mathrm{NW}})}{I_0(\kappa_1 R_{\mathrm{NW}})}+\frac{\varepsilon_2}{\kappa_2}\frac{K_1(\kappa_1 R_{\mathrm{NW}})}{K_0(\kappa_2 R_{\mathrm{NW}})}=0 \tag{1}$$

where $\kappa_i = \sqrt{{k_{\mathrm{spp}}}^2 - \varepsilon_i {k_0}^2}$ ($i$ = 1, 2) represents the transverse wave vector components in the metal core ($i$ = 1) and the surrounding medium ($i$ = 2), with $k_0 = 2\pi/\lambda$ being the free-space wavenumber.[50–52] Here, $I_n$ and $K_n$ are the modified Bessel functions of the first and second kinds of order $n$. Equation (1) was solved numerically using Wolfram Mathematica's FindRoot algorithm. Figures 7(a1) and 7(a2) show the spectra of $\beta(\lambda)$ and $\alpha(\lambda)$ of the NW with $R_{\mathrm{NW}}$ = 10, 35 (fitted with polynomial curves), and 70 nm. Both $\beta(\lambda)$ and $\alpha(\lambda)$ increased with decreasing $R_{\mathrm{NW}}$ owing to the tighter field confinement and considerable ohmic loss, respectively. These spectra are consistent with a previous report.[50] We used $\beta(\lambda)$ and $\alpha(\lambda)$ for $R_{\mathrm{NW}}$ = 35 nm, which is the mean value of $R_{\mathrm{NW}}$,[30] for the following analytical model.

We simplified the system composed of a NP and NW, as illustrated in Fig. 7(b), into two coupled NWs ($NW_1$ and $NW_2$) separated by a barrier that determines the coupling between the two FP resonators by the coefficients of amplitude transmission ($t_c$) and amplitude reflection ($r_c = \sqrt{1 - t_c{}^2}$), as shown in Fig. 7(c). The coefficients of amplitude reflection ($r$), thus, the coefficients of amplitude transmission ($\sqrt{1 - r^2}$), of two NW ends are shown in Fig. 7(b). Note that the model assumes that the barrier does not have scattering and absorption losses, which distract from the FP resonance, as shown in Figs. 4 and 6. Thus, this model is applicable to the spectral region $\lambda$ > 600 nm, which is the spectral tail of the coupled LP resonance maxima.

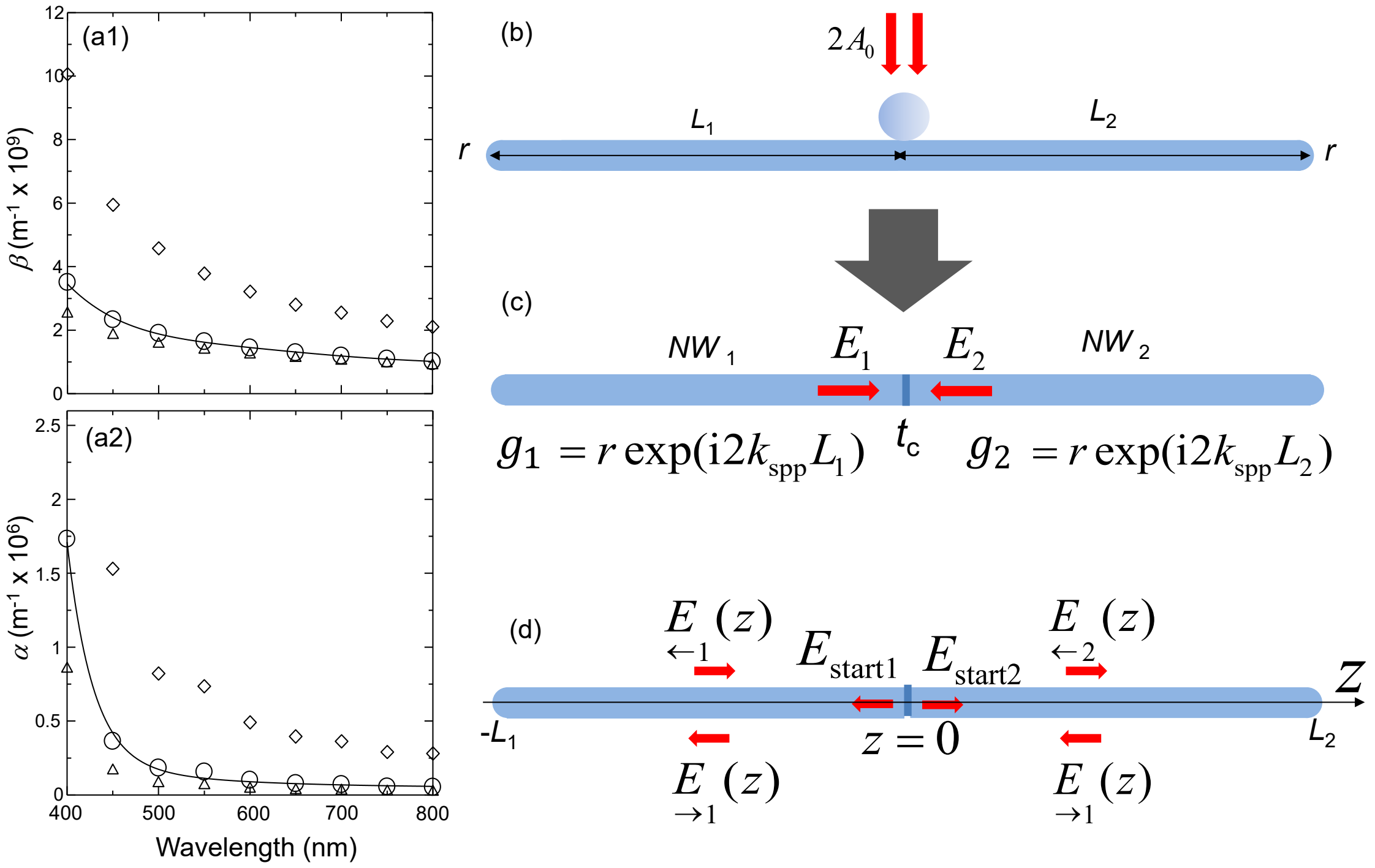

FIG. 7. Spectra of (a1) $\beta(\lambda)$ and (a2) $\alpha(\lambda)$ of an infinitely long NW with $R_{NW}$ of 10 (diamonds), 35 (circles), and 70 nm (triangles). $\beta(\lambda)$ and $\alpha(\lambda)$ for $R_{NW}$ = 35 nm were fitted with a polynomial (solid curves). (b) Schematics of the system composed of a NP and NW excited with an electric field amplitude of $2A_0$ of light, which is equally distributed on both sides of the NP. $r$ represents the coefficient of the amplitude reflection at the NW ends. The definitions of $L_1$ and $L_2$ are identical to those shown in Fig. 5(a). (c) Simplified image of (b) for the analytical model expressed as $NW_1$ and $NW_2$ separated by a barrier whose coefficient of amplitude transmission is $t_c$. $g_1$ and $g_2$ represent the round-trip phase delay and attenuation of the amplitude of the SPPs for $NW_1$ and $NW_2$, respectively. $E_1$ and $E_2$ represent the SPP amplitudes immediately before entering the barrier from the sides of the $NW_1$ and $NW_2$ ends, respectively. (d) Schematic images of propagation of SPPs through z-axis. $\underset{\leftarrow 1}{E}(z)$ ($\underset{\leftarrow 2}{E}(z)$) and $\underset{\rightarrow 1}{E}(z)$ ($\underset{\rightarrow 2}{E}(z)$) are defined as the amplitudes of the SPPs propagating from the barrier $z = 0$ to the left (right) end of $NW_1$ ($NW_2$) and to $z = 0$ from the left (right) end of $NW_1$ ($NW_2$), respectively. $E_{start1}$ and $E_{start2}$ are defined as $E_1$ and $E_2$ just after starting from $z = 0$, respectively.

First, we calculated the radiation spectra from the barrier, as shown in Fig. 7(c), to reproduce the $F_R(\lambda)$ spectra shown in Figs. 4 and 6: White light with an amplitude of $2A_0$ coming from free space irradiates the barrier, and the equally separated light energy $A_0$ and $A_0$ enters $NW_1$ and $NW_2$, respectively. The light energy then propagates as two SPPs along $NW_1$ and $NW_2$. The phase delay and attenuation of the SPP amplitude of one round-trip for $NW_1$ and $NW_2$ are represented by the following two factors:

$$g_1 = r\exp(\mathrm{i}2k_{\mathrm{spp}}L_1) \text{ and} \tag{2}$$

$$g_2 = r\exp(\mathrm{i}2k_{\mathrm{spp}}L_2). \tag{3}$$

The amplitudes of the SPPs immediately before being reflected by the barrier from the sides of $NW_1$ and $NW_2$ are defined as $E_1$ and $E_2$, respectively, as shown in Fig. 7(c). In the two FP resonators, $E_1$ and $E_2$ were assumed to reach steady states after enough number of round trips of the SPPs. Thus, the following self-consistent equations are satisfied:

$$E_1 = (A_0 + r_c E_1 + t_c E_2)g_1 \tag{4}$$

$$\text{and } E_2 = (A_0 + r_c E_2 + t_c E_1)g_2. \tag{5}$$

Note that $A_0 + r_c E_1 + t_c E_2$ in Eq. (4) and $A_0 + r_c E_2 + t_c E_1$ in Eq. (5) indicate the SPPs' field amplitudes just starting at the barrier. Then, $E_1$ and $E_2$ can be derived by solving Eqs. (4) and (5) as

$$E_1 = \frac{A_0 g_1 (1 - r_c g_2 + t_c g_2)}{D} \tag{6}$$

$$\text{and } E_2 = \frac{A_0 g_2 (1 - r_c g_1 + t_c g_1)}{D}, \tag{7}$$

where $D = (1 - r_c g_1)(1 - r_c g_2) - t_c^2 g_1 g_2$. The radiation intensity $|E_{rad}|^2$ of the incoherent light from the barrier is expressed by adding the emission efficiency $\gamma$ of the barrier, as follows:

$$|E_{rad}|^2 = |\gamma E_1|^2 + |\gamma E_2|^2. \tag{8}$$

When $t_c$ is zero, the second term in $D$ disappears and each NW independently exhibits an isolated FP resonance. When $t_c$ is nonzero, the second term in $D$ comes into play; the resonance peaks of $NW_1$ and $NW_2$ interfere with each other, leading to the coexistence of fine and broad structures in both $|\gamma E_1|^2$ and $|\gamma E_2|^2$. Thus, the experimental and calculated $F_R(\lambda)$ in Figs. 4(a), 4(b), and 6(c) can be reproduced by Eq. (8).

Second, we calculated the propagation profiles of the SPPs to reproduce the spatial distributions of $\sqrt{F_R(\lambda)}$ along $NW_1$ and $NW_2$, as shown in Figs. 6(a1)–6(a6). We derived the propagation profile of the SPPs in $NW_1$ and applied this derivation method to $NW_2$. Figure 7(d) illustrates that the SPPs passing through an arbitrary point on the z-axis inside

NW$_1$ consist of a forward SPP $\underset{\leftarrow 1}{E}(z)$ and backward SPP $\underset{\rightarrow 1}{E}(z)$, which are the SPPs propagating from the barrier $z = 0$ to the left end of NW$_1$ and from the left end of NW$_1$ to $z = 0$, respectively. Thus, the total field of the SPPs is expressed as

$$E_1(z) = \underset{\leftarrow 1}{E}(z) + \underset{\rightarrow 1}{E}(z). \quad (9)$$

The SPP field immediately after starting from $z = 0$ was defined as $E_{\text{start1}}$. By using $E_{\text{start1}}$, $\underset{\leftarrow 1}{E}(z)$ and $\underset{\rightarrow 1}{E}(z)$ are described as

$$\underset{\leftarrow 1}{E}(z) = E_{\text{start1}} \exp[-\mathrm{i}k_{\text{spp}}z] \quad (10)$$

$$\text{and } \underset{\rightarrow 1}{E}(z) = E_{\text{start1}} r \exp[\mathrm{i}k_{\text{spp}}(2L_1 + z)]. \quad (11)$$

Note that Eq. (11) indicates that the SPP starts from $z = 0$, reflects at $z = L_1$, and reaches point $z$. Regarding the definition of $E_1$ in Eq. (6), $E_{\text{start1}}$ satisfies the relationship $E_1 = E_{\text{start1}} \times g_1$ under a steady state. Thus, $E_1(z)$ can be expressed as

$$E_1(z) = \frac{A_0(1 - r_c g_2 + t_c g_2)}{D}\{\exp[-\mathrm{i}k_{\text{spp}}z] + r \exp[\mathrm{i}k_{\text{spp}}(2L_1 + z)]\}. \quad (12)$$

$E_2(z)$ can be derived similarly as

$$E_2(z) = \frac{A_0(1 - r_c g_1 + t_c g_1)}{D}\{\exp[\mathrm{i}k_{\text{spp}}z] + r \exp[\mathrm{i}k_{\text{spp}}(2L_2 - z)]\}. \quad (13)$$

It is expected that Eqs. (12) and (13) reproduce the calculated SPP propagation shown in Fig. 6(a).

We evaluated the experimental and FDTD calculations shown in Figs. 4 and 6, respectively, using the analytical model expressed in Eqs. (8), (12), and (13). This model has key parameters such as $r$, $t_c$, $L_1$ and $L_2$. In the experimental and FDTD calculations, both $L_1$ and $L_2$ are known, and the value of $r$ is fixed as $\sqrt{0.27}$.[34] Thus, a comparison between the FDTD and analytical calculations can determine the values of $t_c$. We first evaluated $t_c$ dependence of the radiation intensity $|E_{rad}(\lambda)|^2$ spectra, and then the propagation profiles of $|E_1(z,\lambda)|$ and $|E_2(z,\lambda)|$ for the six conditions of $L_1$ and $L_2$, as shown in Figs. 5(a1)–5(a6).

The $t_c$ dependence of $|E_1(\lambda)|^2$, $|E_2(\lambda)|^2$, and $|E_{rad}(\lambda)|^2$ spectra with $\gamma = 1$ were examined by Eqs. (6)–(8) using six sets of $L_1$ and $L_2$, as illustrated in Figs. 5(a1)–5(a6). Figures 8(a)–8(d) show the $t_c$ dependence of the above three types of spectra. The presence of FP resonance peaks at $\lambda < 600$ nm was due to the lack of attenuation of SPPs by Rayleigh scattering of the coupled LP. When $t_c$ is zero [Figs. 8(a1)–8(a6)], $|E_{rad}(\lambda)|^2$ spectra are simply the superposition of two independent FP resonances with the length of $L_1$ and $L_2$. Thus, $|E_{rad}(\lambda)|^2$ spectra are expressed as the sum of weak fine and strong broad oscillatory structures as the difference between $L_1$ and $L_2$ increases. As $t_c$ increases, both FP resonances begin to interfere destructively and constructively with each other, as shown in Figs. 8(b)–8(d). Thus, $|E_1(\lambda)|^2$ and $|E_2(\lambda)|^2$ exhibit both fine and broad oscillatory structures. In addition, the fine oscillatory structures become more significant as $t_c$ increases [Figs. 8(a6)–8(d6)]. As $t_c$ increases from 0.5 to 1.0 [Figs. 8(b) and 8(d)], the effect of the SPP propagation loss becomes significant, resulting in shallowing oscillatory structures in $|E_{rad}(\lambda)|^2$, $|E_1(\lambda)|^2$, and $|E_2(\lambda)|^2$. In other words, the ohmic loss reduces the total round-trip distance of SPPs, thereby decreasing the interference contrast in FP resonance spectra. This mechanism of shallowing the oscillatory structures was previously discussed for isolated NWs.[34,44,45]

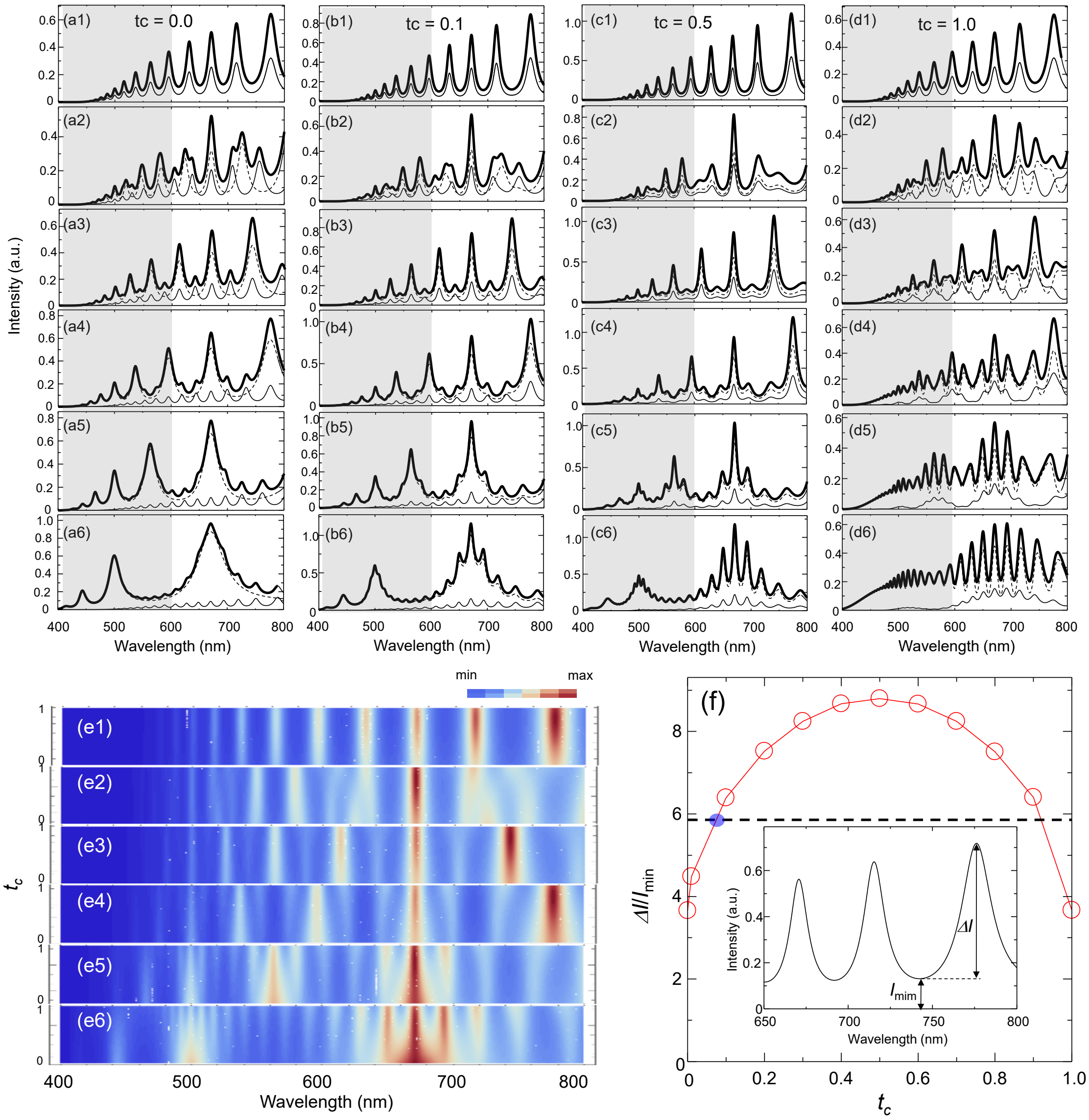

FIG. 8. (a1)–(a6), (b1)–(b6), (c1)–(c6), and (d1)–(d6) $t_c$ dependence of spectra of $|E_{rad}(\lambda)|^2$ (bold solid curves), $|E_1(\lambda)|^2$ (dashed curves), and $|E_2(\lambda)|^2$ (thin solid curves) at the barrier $z = 0$. The values of $t_c$ are indicated in the top panels. The regions <600 nm are sealed by grey panels because FP resonances of these regions disappear by Raleigh scattering of coupled LP resonance at the junctions in the real cases. (e1)–(e6) $t_c$ dependence of $|E_{rad}(\lambda)|^2$ spectrum at the barrier $z$ = 0 expressed as contour maps. (f) $t_c$ dependence of $\Delta I/I_{min}$ under the condition of $L_1 = L_2 = 3$ μm at the barrier $z$ = 0. The inset shows the definitions of $\Delta I$ and $I_{min}$. The values of $L_1$ and $L_2$ are the same as defined in Figs. 5(a1)–5(a6).

Figures 8(e1)–8(e6) show the contour plots of the $|E_{rad}(\lambda)|^2$ spectra with continuously varying $t_c$ from 0 to 1.0 for the six sets of $L_1$ and $L_2$ [Figs. 5(a1)–5(a6)]. With increasing $t_c$, fine oscillatory structures clearly appear in the broad oscillatory structures owing to the interaction between the two FP resonators of $NW_1$ and $NW_2$, supporting the discussion on Figs. 8(a)–8(d). A comparison between the interference contrast of the FDTD calculation and that of the analytical model calculation can be used to derive the $t_c$ in the FDTD calculation. Thus, we defined a factor indicating the degree of interference contrast as $\Delta I/I_{min}$, as defined in the inset of Fig. 8(f) under the condition of $L_1 = L_2 = 3$ μm. We then compared the $\Delta I/I_{min}$ of the analytical method with that of the FDTD calculations. Figure 8(f) shows the increase in $\Delta I/I_{min}$ until $t_c = 0.5$, and then the decrease until $t_c = 1.0$. The increase indicates that the destructive and constructive interferences between the two FP resonances become more significant with increasing $t_c$, whereas the decrease indicates an increase in the ohmic loss due to the increasing effective propagation lengths of the SPPs, overcoming the interference effect appearing in the region of $t_c < 0.5$. Figure 8(f)

shows that $\Delta I/I_{\min}$ of the FDTD calculation is reproduced by $t_c$ around 0.1 and 0.9. Regarding the weak fine oscillatory structures in the $F_R(\lambda)$ spectrum obtained by the FDTD calculation in Fig. 6(c5) and 6(c6), the value of $t_c$ reproducing the FDTD calculation was determined to be 0.1. Therefore, this value was applied to the analytical model to examine the experimental results.

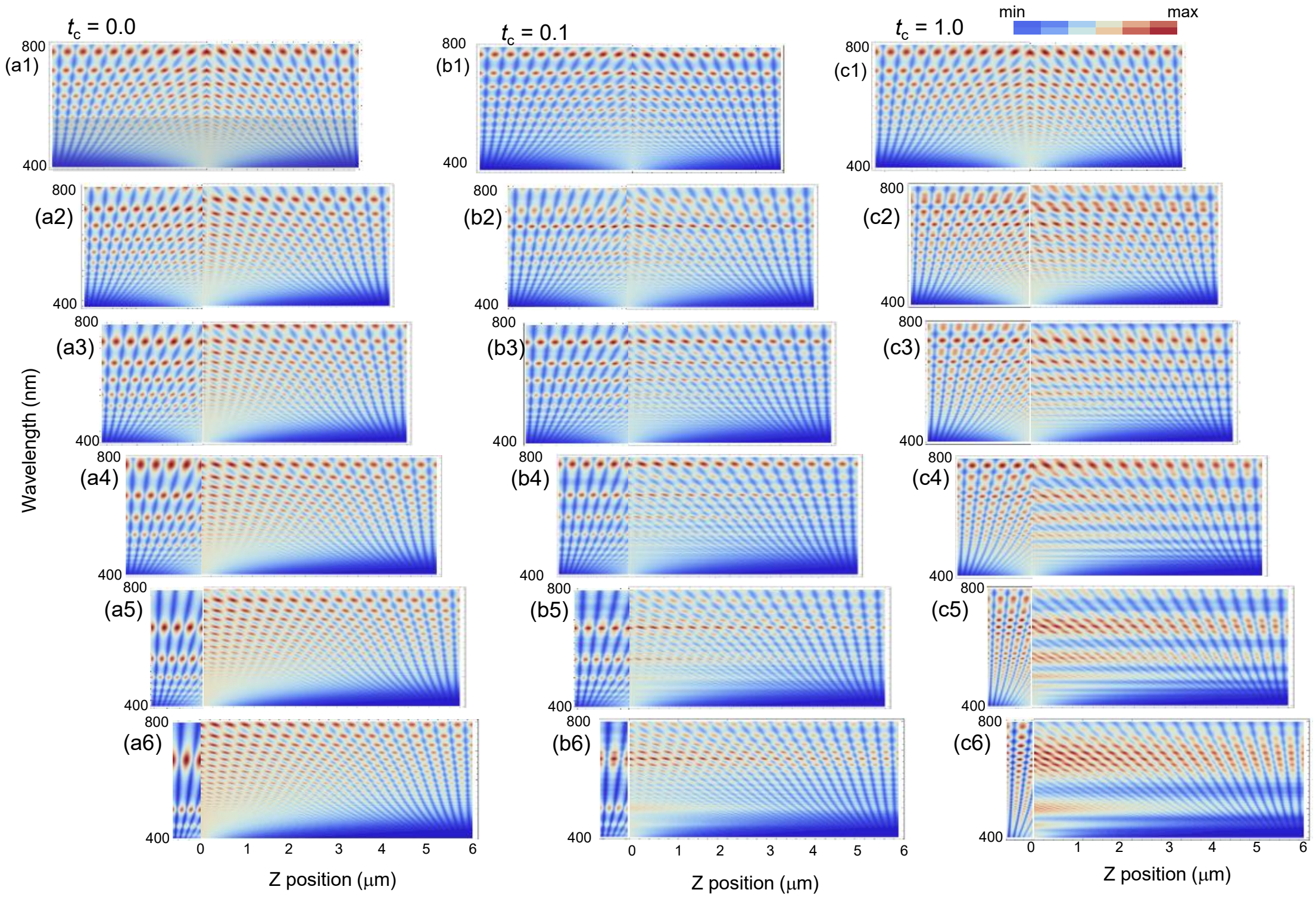

FIG. 9. (a1)–(a6), (b1)–(b6), and (c1)–(c6) $t_c$ dependence of the propagation profiles of $|E_1(z,\lambda)|$ and $|E_2(z,\lambda)|$ expressed as contour maps. The values of $t_c$ are shown in the top panels; the values of $L_1$ and $L_2$ for (a1)–(a6), (b1)–(b6), and (c1)–(c6)are the same as defined in Figs. 5(a1)–5(a6).

The $t_c$ dependence of the SPP propagation profiles was examined using $|E_1(z,\lambda)|$ and $|E_2(z,\lambda)|$ in Eqs. (12) and (13) for six sets of $L_1$ and $L_2$ [Figs. 5(a1)–5(a6)]. Figures 9(a)– 9(c) show the propagation profiles with $t_c$ values of 0, 0.1, and 1.0, respectively. At first glance, all the propagation profiles appear to reproduce the FDTD calculation results shown in Figs. 6(a1)–6(a6) for $\lambda > 600$ nm. The presence of oscillatory structures for $\lambda <$ 600 nm in Figs. 9(a)–9(c) is due to the lack of attenuation of the SPPs by the Rayleigh scattering of the coupled LP. The oscillatory structures disappeared at $\lambda < 400$ nm owing to an increase in $\alpha(\lambda)$, as shown in Fig. 7(a2). For $t_c = 0$, as shown in Figs. 9(a1)–9(a6), the two FP resonators are completely separated from each other, and two independent oscillatory structures along the z-axis are observed on both sides across $z = 0$. As $L_1$ and $L_2$ become shorter and longer, their oscillatory structures in $|E_1(z,\lambda)|$ and $|E_2(z,\lambda)|$ against $\lambda$-axis become broader and finer, respectively, owing to the periodic conditions of the FP resonance expressed in Eqs. (2) and (3), respectively. A careful examination of these SPP propagation profiles reveals that the oscillatory structures of $NW_1$ ($NW_2$) appear in $NW_2$ ($NW_1$) with increasing $t_c$ because $|E_2(z,\lambda)|$ ($|E_1(z,\lambda)|$) penetrates $NW_1$ ($NW_2$). The appearance of the FP interference structures in another NW became clearer with increasing $t_c$ and the difference between $L_1$ and $L_2$. Such an appearance is most clearly observed in Fig. 9(c6) as fine and broad oscillatory structures in $NW_1$ and $NW_2$ regions, respectively. We determined $t_c = 0.1$ by comparing $F_R(\lambda)$ and $|E_{rad}(\lambda)|^2$ based on Figs. 6(c1) and 8(f). The propagation profiles of $|E_1(z,\lambda)|$ and $|E_2(z,\lambda)|$ at $t_c = 0.1$ are also consistent with those of the FDTD calculations shown in Figs. 6(a1)–6(a6). Because the FDTD calculations reproduced the experimental spectral properties well, this value of $t_c$ was applicable to the evaluation of the experimental $\sigma_{sca}(\lambda)$ and $F_R(\lambda)$ spectra.

Finally, we examined the spectral properties of $\sigma_{sca}(\lambda)$ and $F_R(\lambda)$ in Fig. 4 using an analytical model with the $R_{NW}$ and $t_c$ fixed at 35 nm and 0.1, respectively. The values of $L_1$ and $L_2$ were estimated using the dark-field images shown in Fig. 4. Figures 10(a) and 10(b) show the spectra of $|E_{rad}(\lambda)|^2$, $|E_1(\lambda)|^2$, and $|E_2(\lambda)|^2$ for the NWs separated by barriers near the centers and ends of the NWs, respectively. The panels in Figs. 10(a1)–10(a4) and 10(b1)–10(b4) show the results reproduced for Figs. 4(a1)–4(a4) and 4(b1)–4(b4). Note that this model did not consider attenuation by Rayleigh scattering of NPs (Fig. 4), resulting in the presence of oscillatory structures for $\lambda < 600$ nm, which did not appear in either the experimental or FDTD calculation results in Figs. 4 and 6.

For the near center cases [Figs. 10(a1)–10(a4)], the intensities of $|E_1(\lambda)|^2$ and $|E_2(\lambda)|^2$ are comparable because their ohmic losses are close to each other. Smaller differences between $L_1$ and $L_2$ resulted in complex or oscillatory spectral structures in $|E_{rad}(\lambda)|^2$ depending on the $|E_1(\lambda)|^2$ and $|E_2(\lambda)|^2$ spectra. Longer and shorter NWs exhibited broader and finer oscillatory structures, as shown in Figs. 10(a1) and 10(a4). These spectral properties are qualitatively consistent with those shown in Figs. 4(a1)–4(a4). For the near end cases [Figs. 10(b1)–10(b4)], the intensities of $|E_1(\lambda)|^2$ are several times larger than those of $|E_2(\lambda)|^2$ because their ohmic losses differ largely, resulting in the complex structures composed of large broad and small fine peaks in $|E_{rad}(\lambda)|.^2$ These spectral properties for $\lambda > 600$ nm appear qualitatively consistent with both $\sigma_{sca}(\lambda)$ and $F_R(\lambda)$ in Figs. 4(b1)–4(b4). These consistencies indicate that the experimentally observed complex oscillatory spectral structures resulted from the interaction between the two FP resonators separated by a NP on a NW.

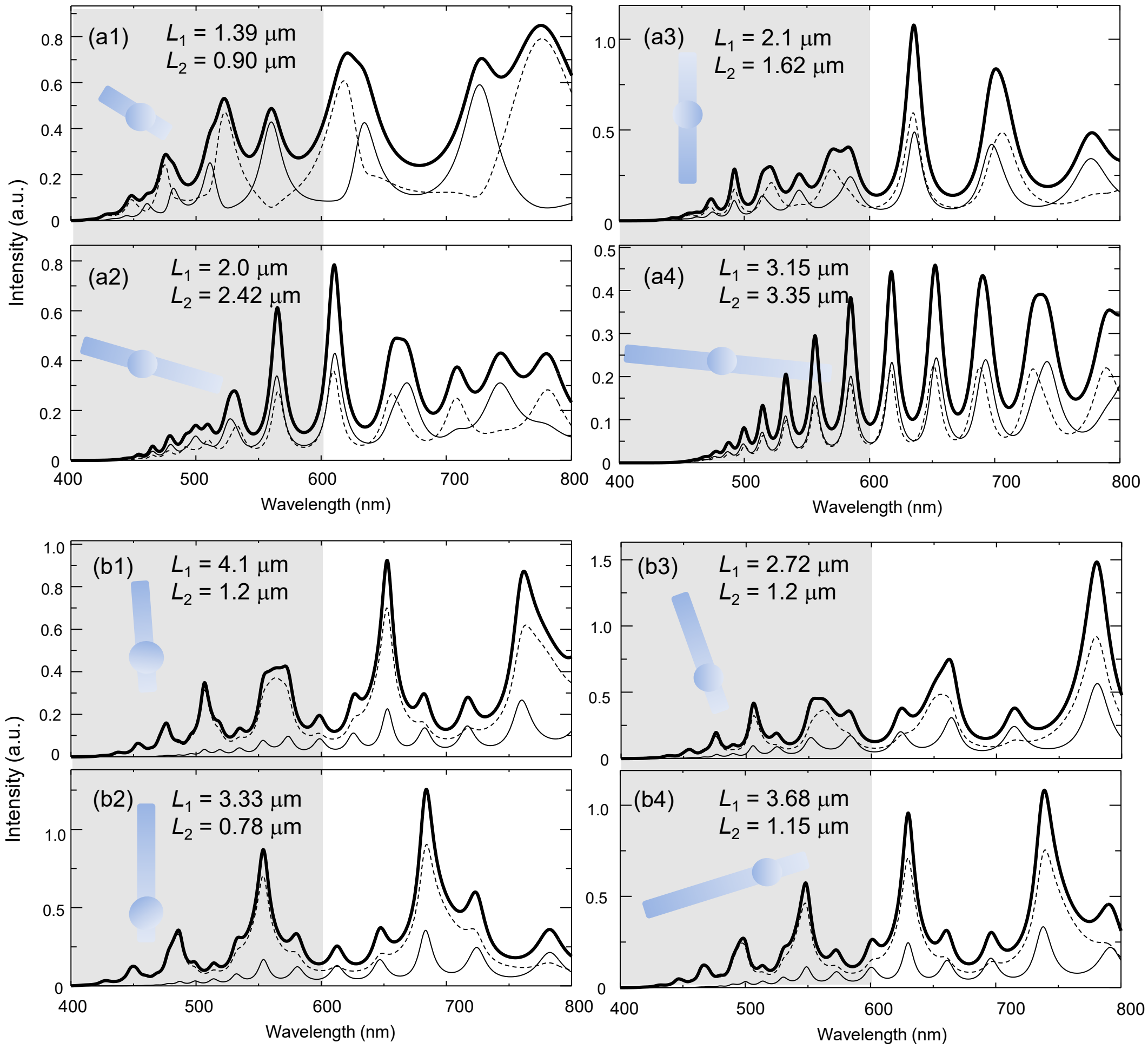

FIG. 10. (a1)–(a4) Schematics and spectra of $|E_{rad}(\lambda)|^2$ (bold solid curves), $|E_1(\lambda)|^2$ (dashed curves),

and $|E_2(\lambda)|^2$ (thin solid curves) at the barrier $z = 0$ reproducing $\sigma_{sca}(\lambda)$ and $F_R(\lambda)$ in Figs. 4(a1)–4(a4). The values of $L_1$ and $L_2$ were estimated using the dark-field images in Figs. 4(a1)–4(a4). (b1)–(b4) Schematics and spectra of $|E_{rad}(\lambda)|^2$ (bold solid curves), $|E_1(\lambda)|^2$ (dashed curves), and $|E_2(\lambda)|^2$ (thin solid curves) at the barrier $z = 0$ reproducing $\sigma_{sca}(\lambda)$ and $F_R(\lambda)$ in Figs. 4(b1)–4(b4). The values of $L_1$ and $L_2$ were estimated using the dark-field images in Figs. 4(b1)–4(b4). The regions <600 nm are sealed by grey panels because FP resonances of these regions disappear by Raleigh scattering of coupled LP resonance at the junctions in the real cases.

We now discuss the deviation between the calculated and experimental results. The free spectral ranges (FSRs), which are the spacing between two successive FP resonance spectral peaks in $\sigma_{sca}(\lambda)$ and $F_R(\lambda)$ in Figs. 10(a) and 10(b), appear wider than those in Figs. 4(a) and 4(b). We consider that this deviation is mainly due to the overestimation of $R_{NW}$ in the analytical model because the value of $\beta(\lambda)$ increases with decreasing $R_{NW}$ [Fig. 7(a1)]. A smaller $R_{NW}$, which results in narrower FSRs, may improve the similarity between the calculated and experimental results. However, the purpose of the calculation was not to precisely reproduce the experimental results but to clarify the mechanism of the experimental spectral properties using known parameters. With decreasing $R_{NW}$, the values of $\alpha(\lambda)$ increased, as shown in Fig. 7(a2). Thus, the values of $\Delta I/I_{min}$ are expected to decrease because of the larger propagation loss of the SPPs. This decrease in $\Delta I/I_{min}$ may be reasonable by comparing Figs. 4(a) and 4(b) with Figs. 10(a) and 10(b), respectively. Such comparisons of FSRs and $\Delta I/I_{min}$ may be useful for estimating the effective values of $R_{NW}$.

## IV. CONCLUSION

In this study, we investigated Rayleigh scattering cross-section $\sigma_{sca}(\lambda)$ and enhancement factor $F_R(\lambda)$ spectra of the HSs between silver NPs and NWs. When a NP is located close to the center of the NW, oscillatory spectral structures appear in $\sigma_{sca}(\lambda)$ and $F_R(\lambda)$ spectra. When a NP is located close to the end of a NW, the spectral structures are composed of large broad and small fine peaks. FDTD calculations of $\sigma_{sca}(\lambda)$ and $F_R(\lambda)$ spectra and the spatial distribution of $F_R(\lambda)$ indicated that the experimental spectral structures originated from the interaction between the two plasmonic FP resonators within the NW separated by the NP. Based on this indication, we developed an analytical model of coupled FP resonators and evaluated their spectral properties using experimentally measured NW lengths and NP positions by varying the inter-resonator coupling efficiency. This analysis revealed that the interference and superposition of the SPPs of the two FP resonators resulted in these spectral features in $\sigma_{sca}(\lambda)$ and $F_R(\lambda)$. Such a method combining SEF spectroscopy and the coupled FP resonator model is useful for evaluating various HSs of SE spectroscopy utilizing NWs,[40,53,54] and is thus expected to contribute to connecting SE spectroscopy and phenomena related to cavity-QED.[12,55,56]

## ACKNOWLEDGMENTS

The authors thank Prof. Jeyadevan Balachandran (University of Shiga Prefecture) for providing the dispersion of silver nanowire. This work was supported by JSPS KAKENHI Grant-in-Aid for Scientific Research (C) (Grant Nos. 25K08520 and 25K08506).

## DATA AVAILABILITY

Data supporting the findings of this study are available from the corresponding authors upon request.